# Taylor & Francis Word Template for journal articles

Pierfrancesco Leonardi[a], Vincenza Torrisi[b] , Andrea Araldo[c], Matteo Ignaccolo[a]

[a]*Department of Civil Engineering and Architecture, University of Catania, Italy; vincenza.torrisi@hotmail.it*

[b]*Department of Electric, Electronic and Computer Engineering, University of Catania, Italy*

[c]*Institut Polytechnique de Paris - Télécom SudParis, France*

*\*vincenza.torrisi@unict.it*

# Data-driven assessment for the predictability of On-Demand Responsive Transit

By adapting bus routes to users' requests, Demand-Responsive Transit (DRT) can serve low-demand areas more efficiently than conventional fixed-line buses. However, a main barrier to its adoption of DRT is its unpredictability, i.e., it is not possible to know a-priori how much time a certain trip will take, especially when no large prebooking is imposed. To remove this barrier, we propose a data-driven method that, based on few previously observed trips, quantifies the level of *predictability* of a DRT service. We simulate different scenarios in VISUM in two Italian cities. We find that, above reasonable levels of flexibility, DRT is more predictable than one would expect, as it is possible to build a model that is able to provide a time indication with more than 90% reliability. We show how our method can support the operators in dimensioning of the service to ensure sufficient predictability.

## 1. Introduction

In low density areas, low ridership forces conventional Public Transport (PT) to provide a low-frequency low-coverage service, to prevent cost-per-passenger from exploding. This results in poor accessibility and car-dependency. DRT is potentially more efficient than conventional PT in such areas and has been proposed as a complement to conventional PT (Calabrò et al., 2023), Wang et al., 2024) or as a replacement (Coutinho et al., 2020). DRT services operate in several territories, but their adoption remains marginal.

From experiences with operating DRT services, a main barrier is the unpredictability of travel times, which cannot be known in advance like in conventional PT, since they depend on detours due to the sharing. Thus, DRT is often perceived as unreliable. Examples of complains from DRT users are "longer travel time from estimated" and

"reliability in comparison to taxi and car use" (Weckström et al., 2018).

When dimensioning the service, DRT planning tools, such as PTV Maas Planner, provide levers for limiting unpredictability, e.g. by imposing a maximum Detour Factor (DF), which is the ratio between the time to go from an origin to a destination via DRT and via the shortest path without detours. Via simulations the tools find the fleet size required to match the maximum DF. However, most of trips are likely to experience much smaller DFs and thus communicating to users the maximum DF as an indication of level of service would be counterproductive. If the maximum DF is too tight, the required fleet size would be too large, hampering financial viability, another barrier for DRT (Davison et al., 2014). This would also reduce occupancy, and thus environmental performance. Therefore, a predictability of DRT cannot be controlled by just planning decisions.

In some cases, trips can be prebooked. This may allow the operator to provide users with a more precise indication of the trip time. However, if prebooking is high, trips cannot be "spontaneous," hindering the attractiveness of the DRT system. We assume in this work prebooking is not needed (or can be in the order of 15 minutes).

A DRT service is predictable if it is possible to provide users with a useful indication of how much time their trip would take if they used DRT in a certain period of day. Such an indication is useful if it allows a user to decide whether it is worth performing such a trip via DRT or not. Such an indication can thus be a time window. Moreover, such an indication must be reliable, i.e., the proposed window should include the actual DRT time a high percentage of times.

In this paper, we devise a model to provide such an indication. Our approach is data-driven and sample-efficient, as the model is constructed based on relatively few previously observed trips. We evaluate our method on several scenarios, simulated in

Visum, representing a DRT system in two different Italian cities. We show that DRT is indeed predictable, as via our data-driven approach we can provide indications that are correct up to 95% of times. However, when DRT is under-dimensioned, not only it degrades in general level of service, but also predictability.

We believe that our method can be used (i) in real DRT deployments, to provide online information to users and encourage their adoption of DRT, (ii) ex-post, as a way to evaluate the quality of a service under the lens of predictability and (iii) ex-ante: if nowadays DRT is dimensioned to provide a certain level of service, in the future it could be dimensioned to reach certain levels of predictability, so as to increase user adoption.

Thus, the remind of the paper is organized as follows: Section 1 provided an introduction on the topic of DRT and highlighted the main aim of the research and its implications, supported by the related works discussed in Section 2; then Section 3 will describe the proposed methodological approach to evaluate the predictability of DRT, by using as input data the numerical results obtained by simulated scenarios related to two real case studies, i.e. the suburb of Partanna Mondello within the city Palermo, and the city of Acireale, in Catania (Italy), described in Section 4. Next, Section 5 will expose the results, demonstrating first the absence of a direct dependence between distance and DRT travel times and consequently, the need to assess predictability through a regression model. Additionally, within this section will be reported the results of further investigations (i.e. to evaluate the service dimensioning and the demand variability) applied to the second real-case study of Acireale, characterized by a different DRT service, to test the replicability of the proposed methodology. Finally, Section 6 will conclude the paper and provide insights for future research development.

## 2. Related work

In the transport literature, the term "predictability" is broadly used to describe the stability, certainty, and reliability of travel conditions. Taylor (2013) and Rasouli & Timmermans (2014) offer a comprehensive review of studies on the predictability of travel times, emphasizing how uncertainty influences travel behaviour. Their focus is on daily fluctuations in travel times and how travellers, equipped with either historical or real-time route information, can mitigate the negative effects of unreliability and uncertainty.

A large corpus of work associates reliability with accessibility, which measures the ease in accessing opportunities, e.g., job, education, healthcare, leisure places. Bimpou and Ferguson (2020) consider the variability in travel times caused by disruptions from anomalous events, e.g. congestion, road construction, and traffic accidents. The authors suggest a travel time reliability metric derived from the upper and lower percentiles of the distribution of observed travel times along a route between a specified origin and destination during a particular time of day and within a defined timeframe. The further employ this reliability metric into a dynamic definition of accessibility.

Bills and Carrel (2021) propose a joint accessibility-reliability measurement. Relying on a large amount of historical Automatic Vehicle Location data, the authors measure the day-to-day travel time variability that must be considered by traveller performing their trips. Specifically, the authors consider the accessibility related to the "total travel time budget" which represents the expected or scheduled travel time plus an additional "buffer time", accounting for the delays. This time budget is associated to different passengers as their behavioural adaptations can vary. However, this approach presents a limitation since within the computation of accessibility, the destinations which do not lie within the travel time budget are a priori excluded.

Lee and Miller (2020) develop a general analytical framework to assess accessibility by taking into account the heterogeneous strategies employed by private car and public transport travellers to deal with travel time uncertainty, e.g. departing earlier to guarantee a safety margin and selecting routes based on more reliable travel times instead of simply the fastest options. They refer to different safety margin plans, represented considering the effective travel time, i.e. the expected time plus the safety margin, founding heterogeneity in routing strategy (i.e. faster travel times vs. higher reliability).

Authors in both works are considering traditional public transport systems, i.e. public transport in San Francisco for Bills and Carrell, (2021) and a new bus rapid service in Columbus, Ohio, USA for Lee and Miller (2020), to whom they apply their methodology. In such cases, for a public transport system it is relatively easy to gather data on travel time fluctuations from day to day or hour to hour, across different weather conditions and other variables. This allows travellers to accumulate experience with these variations, enabling them to form a reasonable understanding of travel time patterns under various conditions, aligning with the objective interpretation of probability. For public transport, the routes of the vehicles are always the same, and therefore it is a good approach to consider the average travel time. On the other hand, for DRT systems this approach cannot work because often the data for averaging are not even available. In fact, DRT systems are not characterised by a graph, and therefore it is sometimes necessary to calculate the travel time required from an origin to a destination even when the route has not been explored. Therefore, we must resort to a supervised learning reliability method.

Considering innovative transport solutions such as on-demand services, where multiple users can travel together in the same vehicle, there is significant potential to enhance mobility and reduce congestion. However, the variability in travel times tends to be greater due to the sharing of the trips, which may result in additional detours that extend waiting times and overall delays, as well as the risk of users being rejected by the system. In this direction, Fielbaum and Alonso-Mora (2020) explore various sources of unreliability associated with ridesharing systems, e.g. different assignment of requests to vehicles and fluctuations in user wait times during a single journey and across multiple instances of the same trip. They propose a routing and assignment method based on a test case in New York City and find that when a user experiences an increase in waiting time, this additional duration is comparable to the average waiting time across the entire system, with similar findings for total delays. A study to increase understanding of the flexibility-uncertainty time inherent in the DRT service is presented by González et al. 2017. Their study shows the switch from fixed line to variable line is perceived by users, proposing a fuzzy variable time inherent to DRT in the mode choice decision, through a Stated

A study related to the public transport prediction was also conducted by Peled ed al. (2021), who analyse the case study of public transport trips in Metropolitan Copenhagen, Denmark. They hypothetically consider "pilot" experiments, with 6 buses that dynamically serve the most active Public Transport stations and measure their performances considering the influence of predictions of transport demand. Similarly, Kun and Hong (2014) adopt a reliability-based formulation for transport network design problem, considering the integration of rapid public transport services (e.g. train and bus) and flexible service, e.g. taxi or dial-a-ride services. The objective of this work is to determine the optimal combination of these two services minimizing the total expected operating costs and serving all realized original destination demands.

However, these approaches are more focused on the demand, which influences the predictability of the service. In our study, we would like to present a second novelty including in the evaluation previous observed DRT trips, thus considering the travel time variability during several days of service and using the results of the predictability evaluation for the dimensioning of the service.

Therefore, to the best of our knowledge, we are the first to propose these research advances: (i) development of a model to assess the predictability associated with an on-demand system with variable routes, stops and timetables and (ii) use of the predictability measure to understand the DRT service configuration.

## 3. Methodology

The simplest indication for users is a prediction of the DRT travel time ($DRTtime$ in short), i.e., the time a user will spend onboard. Due to the inherent stochasticity of DRT routes, it is impossible to predict travel time precisely, which is why DRT is considered inherently unpredictable (Vansteenwegen, et al., 2022). However, users do not really need an exact prediction when deciding whether to use DRT or another mode: for them there is no practical difference between a 25- or 26-minutes trip. A sufficient indication is a time window, within which $DRTtime$ will lie. Such a time window is useful if:

- The window is narrow enough (e.g., a 5-minutes window for a trip 15-minutes trip) and

- The indication is reliable, i.e., the actual *DRTtime* lies into the indicated window most times.

### 3.1 Preliminary definitions

Let us consider trip $i$ and denote with $T_i$ $DRTtime$, which is not known in advance. Trip $i$ is characterized by a vector of features $d_i$, which are static and known in advance. Example of such features are the direct distance between origin and destination, socio-demographic characteristics of the area around the origin and the destination, the centrality of the arcs traversed by the shortest path between origin and destination, etc.. Let us assume that model $f(\cdot)$ is available, which returns, for any trip $i$, a *prediction window* (Equation 1):

$$f(d_i) = [w_i^-, w_i^+] \qquad\qquad (1)$$

We define the **reliability** of model $f(\cdot)$ as the probability that it provides a correct window, i.e., the probability that $[w_i^-, w_i^+] \ni T_i$, for any trip $i$. We define the **predictability** of a DRT service, with respect to model $f(\cdot)$, as the reliability that can be achieved by model $f(\cdot)$, when predicting the travel times experienced in the service.

Note that the predictability of a DRT service is not an absolute measure, but it is relative to model $f(\cdot)$, which must thus be choses appropriately. Indeed, if the model tends to give very large windows (large values of $w_i^+ - w_i^-$), the model will be trivially reliable, but not narrow enough to be of any relevance for the user. By using such a model, we would then improperly conclude that the DRT service is predictable. On the other extreme, if the adopted model $f(\cdot)$ tends to give windows that are too narrow (e.g., 30 seconds), it would be excessively severe to judge the predictability of a DRT service based on such a model.

In what follows we describe how we build model $f(\cdot)$ to quantify the predictability of DRT in this paper. However, we pinpoint that the concepts we present are applicable with other more advanced models $f(\cdot)$, which are beyond the scope of this work.

Note thatthe more the features we consider in $d_i$, the more the expressive power of model $f(\cdot)$. In this paper we show that we already achieve a highly reliable prediction window byincluding in features $d_i$ only the direct distance, i.e., the length of the shortest path (in Km) between origin and destination. In what follows, $d_i$ will thus simply denote the direct distance of trip $i$.

We construct $f(\cdot)$ as a composition of two models. First, we build a regression model $r(\cdot)$, which gives a prediction $\hat{T}_i = r(d_i)$ of $DRT\,time$. Then, we build a window around $\hat{T}_i$, via functions $w^-(\cdot)$, $w^+(\cdot)$. The window related to trip $i$ is (Equation 2):

$$f(d_i) = [w_i^-, w_i^+] = [w^-(\widehat{T_i}), w^+(\hat{T}_i)] = [w^-(r(d_i)), w^+(r(d_i))] \qquad (2)$$

### 3.2. Building regression model $r(\cdot)$

We fit $r(\cdot)$ on training set $D_{tr}$ of previously observed trips. Function $r(\cdot)$ is parametrized by certain coefficients. Training $r(\cdot)$ means to find the values of such coefficients that minimize the mean square error, which is reported in Equation 3:

$$MSE_{tr} = \frac{1}{|D_{tr}|}\Sigma_{i \in D_{tr}}\left(\hat{T}_i - T_i\right)^2 = \frac{1}{|D_{tr}|}\Sigma_{i \in \Omega tr}\left(r(d_i) - T_i\right)^2 \quad (3)$$

We tried several regression model types, such as simple Average, Kriging (with Multiscaling to embed origin and destination positions into the 2-dimensional space– more details in Diepolder et al., 2023), Linear, Polynomial, Exponential and Logarithmic regression. We only report results for Average, Kriging and Linear Regression as all other models showed similar trends.

### 3.3. Building the window

It is natural to fix a window as a fraction $x$ of the estimated travel time, i.e.:

Proportional window:   $w^-(\hat{T}_i) = (1 - x) \cdot \hat{T}_i$ ;   $w^+(\hat{T}_i) = (1 + x) \cdot \hat{T}_i$  (4)

However, for short trips, such a window would be narrower than practically needed and the indication would be unreliable. To avoid this, we define a "stretched window" (Equation 5):

$$w^-(\hat{T}_i) = \begin{cases} \hat{T}_i - W_{min} & if\ \hat{T}_i \le T_{th} \\ (1 - x) \cdot \hat{T}_i & otherwise \end{cases} ; w^+(\hat{T}_i) = \begin{cases} \hat{T}_i + W_{min} & if\ \hat{T}_i \le T_{th} \\ (1 + x) \cdot \hat{T}_i & otherwise \end{cases} \qquad (5)$$

In other words, for trips shorter than $T_{th}$, we will provide a window of length $2W_{min}$. Hyperparameters $W_{min}$ and $T_{th}$ need to be adapted to the scenario under study. For instance, in a scenario with an average DRTtime of about $\overline{T} = 10$ minutes, it is reasonable to assume that, for all trips with predicted DRTtime $\hat{T}_i < 15$ minutes, an error of 5 minutes would be almost acceptable for the user. In this case, we set $T_{th} = 15$ minutes and $W_{min} = 5$ minutes. It is reasonable to assume that the level of indifference increases with the average $\overline{T}$ experience DRTtime. We accordingly construct windows parameters $T_{th}$ and $W_{min}$ to keep ratios $T_{th}/\overline{T} \approx 15/10$ and $W_{min}/T_{th} \approx 5/15$.

### 3.4. Accounting for demand stochasticity when evaluating predictability

The quantification of the predictability of a DRT service, as defined in Sec. [aa-sec-prel-def] is complicated by the stochastic nature of the DRT service. Indeed, even if keep unvaried the operational parameters of the DRT service, the sequence of user requests will not be exactly the same every day. As a consequence, the DRT routes change from a day to another, as well as the consequent experienced DRTtimes.

Let us assume DRT operational characteristics have been fixed a-priori: there is a certain fleet of DRT vehicles, that all start from pre-fixed positions and a certain algorithm is applied to route DRT vehicles. Let $ud$ be the stochastic process representing the user demand from 0:00 to 24:00. Let $ud_\omega$ be one realization of such stochastic process, i.e., a sequence of user requests over 24h, each characterized by the time of request, the requested pick-up time, the origin and the destination. Realizations $ud_\omega$ are indexed by sample $\omega$ in some probability sample space $\Omega$. Sample $\omega$ can be interpreted as a day of DRT operation. We assume requests are generated from a same underlying stochastic process $ud$, in the sense that realizations $ud_\omega$ have all the same statistical properties, in terms of expected request rate and probability distribution of origins and destinations. Such stochastic process $ud$ is generally non-stationary, as demand intensity and regions of interests usually change over the day. Despite this statistical consistence, it is important to note that sequence $ud_\omega$ of requests might be different from day to day.

Based on Sec. 3.1, given any prediction window model $f(\cdot)$, the predictability of a DRT service with respect to $f(\cdot)$ is

$$Pred = \mathbb{P}(T_i \in f(d_i) | i \sim du),$$

where $i \sim du$ indicates that trip $i$ is generated by stochastic process $du$. While this definition is rather abstract, we can approximate it as the fraction of trips of a certain realization $du_\omega$ (i.e., of a certain day) for which the prediction window is correct:

$$\widetilde{Pred}_\omega = \frac{1}{|du_\omega|} \sum_{i \in d_\omega} \mathbb{I}\left(T_i \in f(d_i)\right), \qquad (6)$$

where $\mathbb{I}$ denotes the indicator function, which is $1$ if $T_i$ is contained into the prediction window, $0$ otherwise. It is evident that this approximation of the predictability may change from a day to another. If we can observe multiple days $\Omega_{\text{test}}$ of operation, we can refine such an approximation as

$$\widetilde{Pred} = \frac{1}{|\Omega_{\text{test}}|} \sum_{\omega \in \Omega_{\text{test}}} \widetilde{Pred}_\omega, \qquad (6-bis)$$

So far in this subsection we have tackled the stochasticity of the predictability measure over different days of operation, given a pre-fixed prediction window model $f(\cdot)$. However, we need to also consider that the prediction window model $f(\cdot)$ is data-driven, and is thus itself of stochastic nature. Indeed, it contains regression model $r(\cdot)$ whose parameters depend on the days $\omega \in \Omega_{tr} \subseteq \Omega$ used to train it.

To tackle ~~achieve~~ this, we simulate multiple observation days by stochastically assigning an hourly Origin-Destination (OD) matrix as a base. This matrix represents movements between origin and destination zones. To account for spatial and temporal variability, the trips in the OD matrix for each zone are randomly distributed among the nodes within that zone. By repeating this procedure across several simulations, we capture consistent variability both spatially (i.e. by assigning the demand to different nodes within the zone across iterations) and temporally (i.e. by assigning nodes to varying time intervals within the specified period, such as one hour). Following the transition from a zonal to a nodal matrix for both origin and destination nodes, travel requests are generated. This process enables the creation of a set of travel requests corresponding to different operational days of the simulated DRT service, providing the basis for evaluating the system's predictability.

### 3.5. DRT service model and simulation

We evaluate the proposed methods on data from simulations in VISUM. We use the simulation to generate a set of observations for the analysed DRT service, which allows us to test our methodology and, thus, assess the accuracy of the prediction window. Moreover, through simulation we are able to assess several scenarios under different service operating configurations, considering the performance in terms of percentage of satisfied requests and predictability, thus supporting the operators for dimensioning the service.

We set the sub-network, i.e. the set of links where the DRT vehicles can run; we identify the DRT stops as the potential locations for Pick-ups and Drop-offs (PUDOs). Then, trip requests result from the disaggregation of an Origin-destination matrix. Since origins and destination may not coincide with PUDOs, users might need to walk.

During simulation, "Dispatcher" assigns requests to the closest available vehicle favoring the sharing. When a user requests a trip, the dispatcher provides the $DRT\,time$ that such a trip would take. The user can accept or not the trip. Let us define the $Ideal\,Travel\,Time\,(ITT)$ of a trip as the travel time the trip would take without detours.

The difference between the $DRTtime$ and $ITT$ is the $Detour\ Time$ ($DT$). The system is parametrized by:

- All accepted detour time ($AllAcc.DT$): if $DT < AllAccDT$, a user always accepts the trip;

- Maximum detour time ($MaxDT$): if $DT > MaxDT$, the dispatcher assigns to the user an empty vehicle if available;

- Maximum detour factor ($MaxDF$): if $AllAccDT < DT < MaxDT$ we need to calculate the $DF$. If $DF > MaxDF$ a user never accepts the trip.

We also impose a maximum waiting time at the DRT stop ($MaxWait$) and a maximum walking time ($MaxWalk$). Users can make a pre-booking, stochastically associated to each trip, which limits WaitTime.

## 4. Case studies

We now evaluate our approach and in particular the ability to provide a reliable prediction window. We also show our approach can be used to give an indication of the predictability of a DRT service and how such a predictability indication can guide the dimensioning of a DRT service. We present two case studies characterized by different transport contexts.

### *4.1. Scenarios for Palermo DRT*

The study area includes the neighborhoods of Tommaso Natale and Partanna Mondello, inside Palermo Metropolitan Area (Italy) (Capodici ed al., 2022). PT only consists of 7 bus lines, with low frequencies. For these reasons, the transport supply has been improved by implementing a pilot of a DRT service with nine-seat minivans ran in 2022. The simulation setup described here was initially developed for dimensioning the pilot.

The pilot service implemented in Partanna Mondello is partially dynamic, because it is characterized by predefined route with possible deviations, that can be activated according to the trip requests (see Figure 1). The OD matrix is estimated via a SP survey (Capodici et al., 2022) and all simulated trip requests are with a single passenger and referred to peak hour, from 7:15 a.m. to 8:15 a.m.

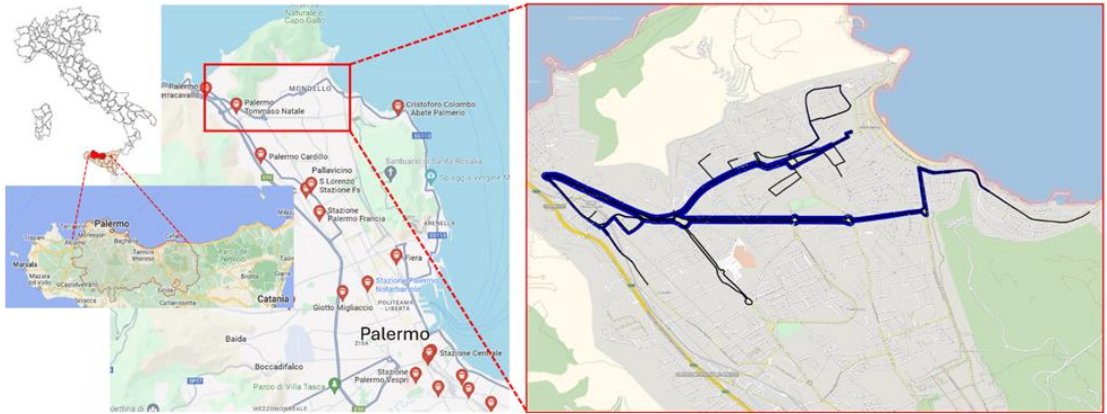

Figure 1 Sub-network for DRT pilot in Partanna Mondello, Palermo (Italy). Source: Authors elaboration from Google Maps and PTV Visum

We simulate five scenarios (Table 1), ordered from the most flexible (i.e., best service to users) to the least.

Table 1 Considered scenarios for Palermo DRT (grey values are simulation output). The meaning of the columns is explained in Sec. 3.4.

| Scenario | MaxDF | MaxDT | All.Acc.DT | Max wait | Max walk | Trip Requests | Satisfied Requests | N. Vehicles | DF |
|---|---|---|---|---|---|---|---|---|---|
| Pal_1 | 2.5 | 20 | 5 | 7 | 7 | 398 | 91% | 42 | 1.58 |
| Pal_2 | 2.5 | 25 | 10 | 7 | 7 | 401 | 91% | 31 | 2.15 |
| Pal_3 | 2.5 | 30 | 15 | 7 | 7 | 401 | 91% | 29 | 2.55 |
| Pal_4 | 5 | 30 | 25 | 10 | 10 | 422 | 97% | 28 | 3.17 |
| Pal_4-subnet | 5 | 30 | 25 | 10 | 10 | 425 | 96% | 23 | 3.31 |

As expected, the first scenarios have higher cost, in terms of vehicles to operate. In all scenarios DRT buses can travel within a sub-network of 995 links, all accessible to cars, except for Scenario 4-subnet: its parameters are like in Scenario 4 but DRT buses can only visit 741 links in which conventional PT is operating. Scenario 4-subnet is envisaged by the PT operator for liability reasons (e.g. restrictions to authorized routes to ensure passenger safety and regulatory compliance). For each scenario we build the model to provide DRTtime indication as described in Section 3.3. Hyperparameters Tth and Wmin, whose values are reported in Table 2, are set according to Sec. 3.2:

Table 2 Parameters of indication window

|  | Sc.1 | Sc.2 | Sc.3 | Sc.4 | Sc.4-subnet |
|---|---|---|---|---|---|
| Average DRTtime[min] | 8 | 10 | 12 | 15 | 17 |
| $T_{th}$[min] | 12 | 15 | 18 | 22 | 25 |
| $W_{min}$[min] | 4 | 5 | 6.5 | 8 | 9 |

## *4.2. Scenarios for Acireale DRT*

We also test our methodology in another real case study in Acireale (Italy), also characterized by weak and spatially dispersed demand. We derive the OD matrix from census data. Most of the considered trips are occasional (i.e. not systematic), to which DRT is particularly suitable. The existing conventional bus service has low frequencies between 60 and 90 minutes. We simulated a DRT service replacing the existing bus service, assuming all bus users would use DRT. The simulated service for Acireale is totally dynamic, with flexible routes and without timetables, and runs on a larger sub-network than Palermo one, with 2818 links and 899 stops (physical and virtual), for the majority virtual stop points (see Figure 2).

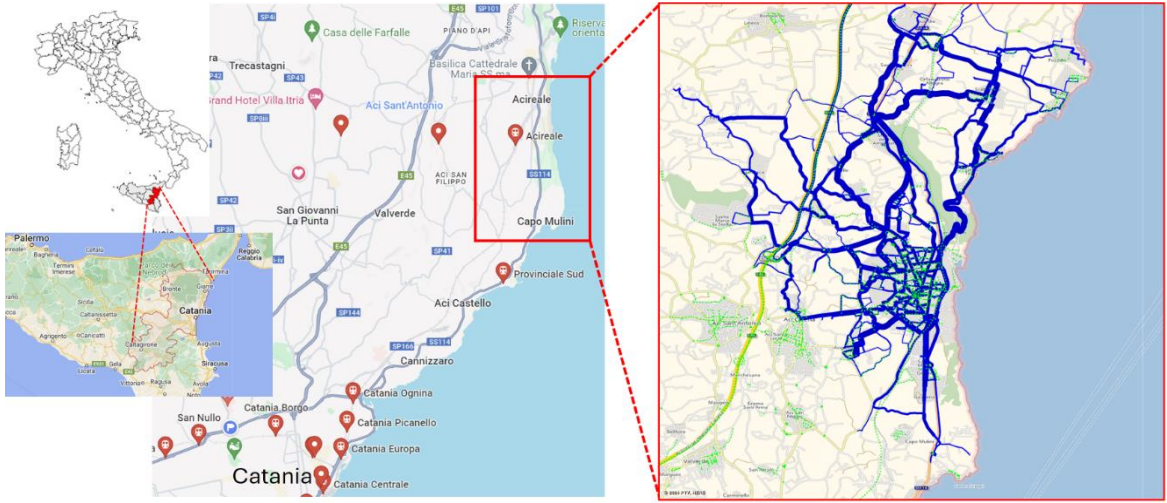

Figure 2 Sub-network for DRT pilot in Acireale, Catania (Italy)

Operational parameters related to the implemented and analyzed scenarios in Acireale are summarized in Table 3.

Table 3: Operational parameters for Acireale DRT scenarios

| Scenario | MaxDF | MaxDT | All.Acc.DT | Max wait | Max walk | Trip Requests | Satisfied Requests | N. Vehicles | DF |
|---|---|---|---|---|---|---|---|---|---|
| 1_ACI | 3.5 | 15 | 8 | 15 | 15 | 201 | 87% | 16 | 3.02 |
| 2_ACI | 2.5 | 12 | 6 | 15 | 15 | 179 | 77% | 16 | 2.55 |
| 3_ACI | 2 | 10 | 6 | 10 | 6 | 174 | 75% | 16 | 2.44 |

### 4.3. Dependency between DRTtime and DirectDistance

We aim to build a DRTtime indication based on features of the trip that are easily known in advance. We exploit the simplest case of DirectDistance $d_i$ of trip $i$, i.e the shortest path between the origin and the destination on the entire graph. Therefore, the reliability of model $f(d_i)$ presented here can be seen as a lower bound of what is achievable if more attributes are considered. Before building the model, we first visually check if the DirectDistance is expressive enough for building an indication of DRTtime.

Figure **3** shows a clear correlation between $d_i$ and $\hat{T}_i$, which encourages on the possibility

to construct a sufficiently reliable prediction window. This suggests that systems we are considering can show a good predictability (which we will analyse more in depth in what follows). Note that in scenario **Pal**-4-subnet, for given values of $d_i$, DRTtime $\hat{T}_i$ can vary a lot. This variability is due to undermining of Scenario 4 and the high circuity of the subnetwork of Scenario 4-subnet, which makes every detour extremely **different from the direct distance.**

This means that a prediction model trying to guess exactly travel time would have bad performance. We will show instead that, despite the variability of DRTtimes, it is possible to build a prediction window (1) that allows to provide reliable indications to users.

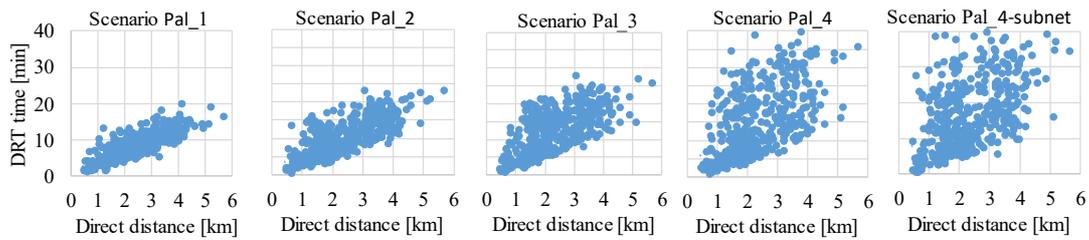

Figure 3 Dependency between DRTtime and DirectDistance for Palermo DRT

We also performed this correlation analysis for the scenarios of Acireale DRT, as shown in Figure 4.

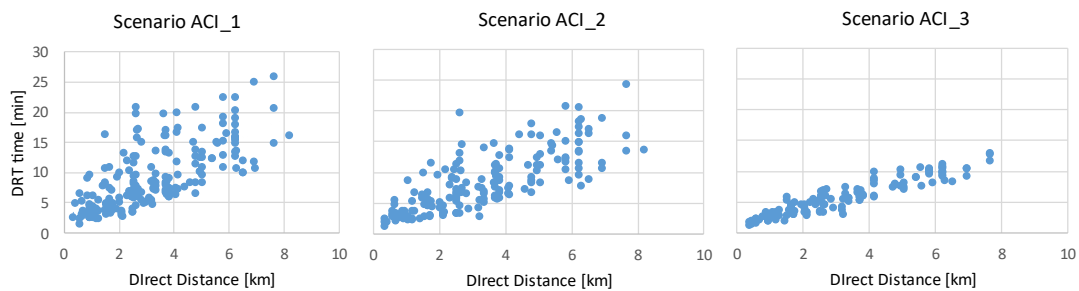

Figure 4 Dependency between DRTtime and DirectDistance

In this case, Direct Distance is expressive as an indication of *DRTtime* in Scenario ACI_3,

which is more constrained service operating parameters, resulting in better service efficiency (i.e. lower DF) at the expense of satisfied requests, which slightly decrease in this last scenario. While Scenario ACI_1 and Scenario ACI_2 are configurations with less restrictive operating parameters enable greater travel sharing, which in turn leads to increased time variability.

## 5. Results and discussion

### 5.1. Choice of regression model

One might be tempted to choose the type of regression model $r(\cdot)$ based on the mean square error (such error is the difference between the predicted time $\hat{T}_i = r(d_i)$ and the true time $T_i$) on the test set, as usually done in data science. However, this would give an indication of the absolute precision of the regression, while we are interested in the reliability of an indication, provided in the form of a window. We thus need another evaluation strategy for which we introduce the reliability curves, as in Figure 3, where we plot the reliability of our indication (2) when changing window narrowness $x$ (see (4)-(5)). Figure 5 is obtained in Scenario Pal_1 (we observe similar trends in the others). To avoid biases, by randomly shuffling and splitting the dataset of all trips, we obtain 10 pairs of training/test sets, with proportions 70%/30%. For each pair, (i) we train the regression model in the training set, (ii) we construct the window as in Section 2 and (iii) for each narrowness x, we measure the reliability on the test set, the fraction of times the actual DRTtime fell into the indicated window, resulting in a reliability curve. The curve with stronger color is the average of the 10 reliability curves.

We observe that: (i) Having a stretched window is crucial to give a reliable indication; (ii) Building a reliable *DRTtime* indication is not trivial (a simple average performs very poorly); (iii) Linear Regression performs best for any window narrowness x. In the rest of the paper, we will thus only use Linear Regression.

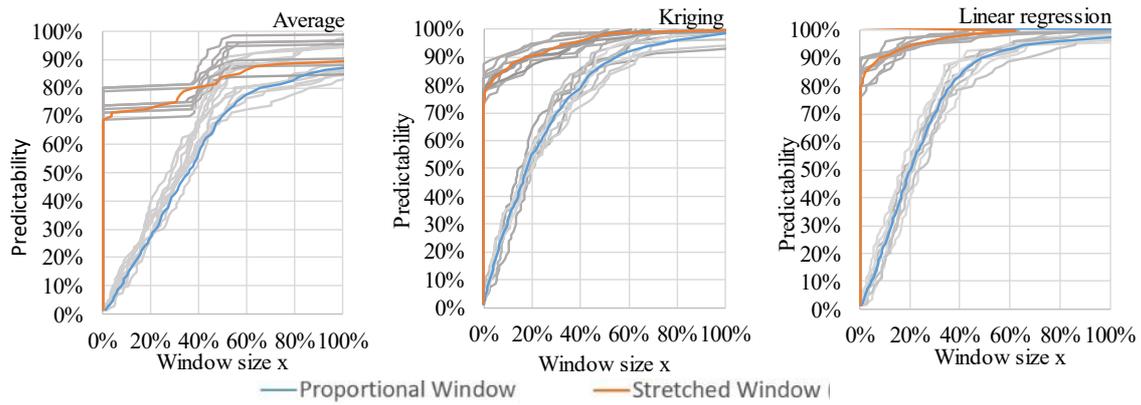

Figure 5 Reliability curves with different regression models

## 5.2. Assessment of the predictability for Palermo DRT

Figure 6 shows the average reliability curves obtained in all scenarios of DRT in Palermo.

We define the predictability of a DRT service as the reliability of DRT indication obtained when window narrowness x=25%. Scenarios Pal_1, Pal_2 and Pal_3 show a predictability between 85% and 95%; while predictability of Scenarios Pal_4 and Pal_4-subnet would not be acceptable for users. Observe that using window stretch parameters $T_{th}$ and $W_{min}$ other than those in Table 2, we would observe slightly different predictability values. However, we do not expect significantly different trends, also considering that the interval of values of $T_{th}$ and $W_{min}$ for which the window stretch is acceptable for users is quite limited.

A DRT operator may dimension the fleet size to achieve a certain level of predictability. Figure 7 compares the scenarios of Table 1. We observe that the more flexible the system is, the more predictable it is, at the price of increasing the fleet. Operators might consider, however, that increasing the fleet not only increases flexibility, but also predictability, thus potentially attracting more users.

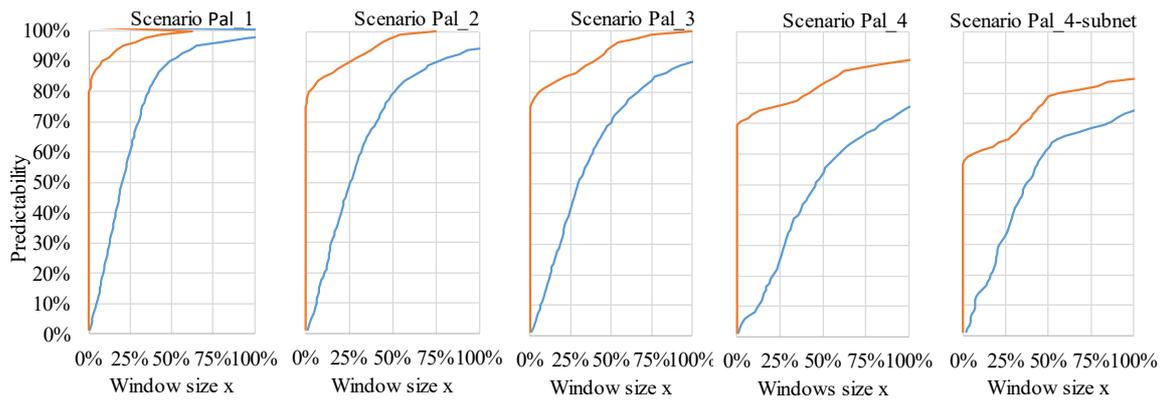

Figure 6 Reliability curves with linear regression

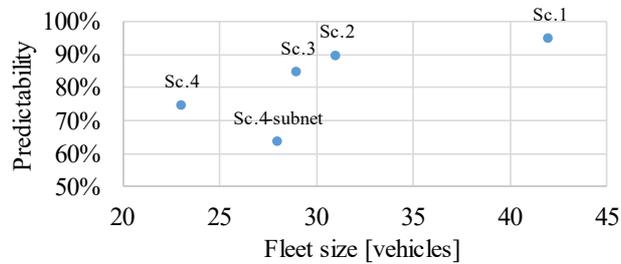

Figure 7 Predictability

### 5.3. Dimensioning Acireale DRT to ensure predictability

To assess the predictability for Acireale and use it as a measure to evaluate the DRT configuration, we performed additional scenarios as a variation of the previous main Scenario ACI_1; Scenario ACI_2 and Scenario ACI_3. Specifically, in this first step of this analysis, we performed an analysis by maintaining fixed the Maximum DF, Maximum DT and All Accepted DT within each main scenario and then varying the maximum walking and waiting time, for a total of 13 simulated scenarios.

The results (Table 4) obtained for Acireale confirmed the trends observed in Palermo. By increasing the service performance, by reducing the maximum DF and the maximum DT (i.e. from Scenario ACI_1 to Scenario ACI_2), there is an improvement in predictability (i.e. from 82% to 86% considering an error window of 25% in terms of estimated DRT times) at the expense of serving 10% fewer of the travel requests (from 87% to 77%). To increase these percentages, operating parameters are further improved in Scenario ACI_3, reaching a predictability of 95%. Nevertheless, the number of satisfied requests became lower (i.e. 75%).

Observing the intermediate scenarios, we identified the best configurations in terms of balancing satisfied requests and predictability, i.e. Scenario ACI_2.3 and Scenario ACI_3.

Table 4 Operational parameters for Acireale DRT additional scenarios and predictability evaluations

| Scenario | MaxDF | MaxDT | All.Acc.DT | Max wait | Max walk | Trip Requests | Satisfied Requests | Predictability [x=25%] | N. Vehicles | DF |
|----------|-------|-------|------------|----------|----------|---------------|--------------------|------------------------|-------------|------|
| ACI_1 | 3.5 | 15 | 8 | 15 | 15 | 201 | 87% | 82% | 16 | 3.02 |
| ACI_1.1 | 3.5 | 15 | 8 | 12 | 12 | 214 | 92% | 81% | 16 | 3.06 |
| ACI_1.2 | 3.5 | 15 | 8 | 12 | 8 | 214 | 92% | 81% | 16 | 3.06 |
| ACI_1.3 | 3.5 | 15 | 8 | 12 | 6 | 211 | 91% | 83% | 16 | 3.08 |
| ACI_1.4 | 3.5 | 15 | 8 | 10 | 6 | 204 | 88% | 84% | 16 | 2.99 |
| ACI_2 | 2.5 | 12 | 6 | 15 | 15 | 179 | 77% | 86% | 16 | 2.55 |
| ACI_2.1 | 2.5 | 12 | 6 | 12 | 8 | 179 | 77% | 86% | 16 | 2.55 |
| ACI_2.2 | 2.5 | 12 | 6 | 12 | 6 | 181 | 78% | 88% | 16 | 2.58 |
| ACI_2.3 | 2.5 | 12 | 6 | 10 | 6 | 190 | 81% | 87% | 16 | 2.52 |
| ACI_3.1 | 2 | 10 | 6 | 15 | 15 | 176 | 76% | 93% | 16 | 2.41 |
| ACI_3.2 | 2 | 10 | 6 | 12 | 8 | 176 | 76% | 93% | 16 | 2.41 |
| ACI_3.3 | 2 | 10 | 6 | 12 | 6 | 174 | 75% | 90% | 16 | 2.44 |
| ACI_3 | 2 | 10 | 6 | 10 | 6 | 174 | 75% | 95% | 16 | 2.44 |

Then, we conducted an analysis by varying the number of vehicles in the fleet to assess whether an increase in fleet size would enhance service performance. The results (Table 5) indicated that an increase of only four vehicles (from 16 to 20 vehicles) in Scenario ACI_2.3c scenario yielded substantial improvements in both the percentage of satisfied requests and predictability, reaching 94% and 92%, respectively.

Table 5 Operational parameters for Acireale DRT additional scenarios and predictability evaluations

| Scenario | MaxDF | MaxDT | All.Acc.DT | Max wait | Max walk | Trip Requests | Satisfied Requests | Predictability [x=25%] | N. Vehicles | DF |
|---|---|---|---|---|---|---|---|---|---|---|
| Aci_2.3a | 2.5 | 12 | 6 | 10 | 6 | 190 | 81% | 87% | **16** | 2.52 |
| Aci_2.3b | 2.5 | 12 | 6 | 10 | 6 | 203 | 88% | 86% | **18** | 2.52 |
| Aci_2.3c | 2.5 | 12 | 6 | 10 | 6 | 217 | 94% | 92% | **20** | 2.51 |
| Aci_3a | 2 | 10 | 6 | 10 | 6 | 174 | 75% | 95% | **16** | 2.44 |
| Aci_3b | 2 | 10 | 6 | 10 | 6 | 189 | 81% | 90% | **18** | 2.44 |
| Aci_3c | 2 | 10 | 6 | 10 | 6 | 217 | 94% | 90% | **20** | 2.33 |

At this point, after choosing the operating parameters and fleet size, the daily variability of demand was assessed. Using a random seed for each simulation, the temporal and spatial demand variability was explored under equal service operating conditions. Specifically, using the same OD matrix and related time series (i.e. flow distribution during each hourly interval), by varying the random seed, 10 lists of travel requests were generated. Each of these lists was used to initiate the dispatching procedure in order to measure satisfied requests and predictability, considering the same operational parameters chosen following fleet sizing, as summarized in Table 6.

Table 6 Selected operational parameters for Acireale DRT

| MaxDF | MaxDT | All.Acc.DT | Max wait | Max walk | N. Vehicles |
|---|---|---|---|---|---|
| 2.5 | 12 | 6 | 10 | 6 | 20 |

The simulation results of the DRT service in the 10 scenarios are as follows (Table 7):

Table 7 Scenario with demand variability for Acireale DRT

| Scenario | Trip Requests | Satisfied Requests | DF |
|---|---|---|---|
| Aci_2.3c.1 | 217 | 94% | 2.51 |
| Aci_2.3c.2 | 219 | 93% | 2.5 |
| Aci_2.3c.3 | 208 | 91% | 2.5 |
| Aci_2.3c.4 | 224 | 97% | 2.34 |
| Aci_2.3c.5 | 216 | 93% | 2.44 |
| Aci_2.3c.6 | 222 | 95% | 2.45 |
| Aci_2.3c.7 | 226 | 98% | 2.49 |
| Aci_2.3c.8 | 223 | 94% | 2.42 |
| Aci_2.3c.9 | 236 | 88% | 2.47 |
| Aci_2.3c.10 | 225 | 96% | 2.44 |

In this case, predictability was not calculated for each individual scenario, but a dataset was used containing all the observations made for the 10 simulated days (the complete list of demands fulfilled in all 10 simulations). From the set of observations (distinguished by day), 10 linear regressions were performed by varying the training set. Specifically, for each linear regression, 2 days (the observations of 2 days/scenarios) were used as the test set and 8 days as the training set.

Table 8 Scenario with demand variability for Acireale DRT

| RUN | Predictability [x=25%] |
|---|---|
| 1 | 89% |
| 2 | 90% |
| 3 | 89% |
| 4 | 86% |
| 5 | 86% |
| 6 | 87% |
| 7 | 86% |
| 8 | 88% |
| 9 | 87% |
| 10 | 87% |

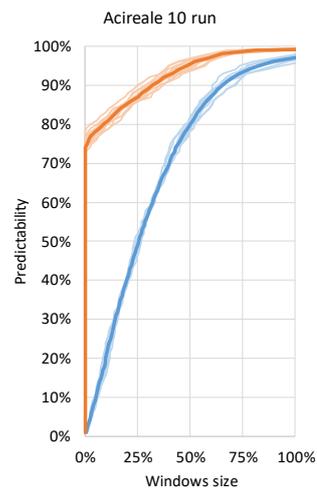

Figure 8 Predictability with demand variability

The results of this service analysis on 10 different days showed a predictability of 87% and a percentage of satisfied users of 94%, averaging over 10 different simulations. We then demonstrated how to size the service in terms of fleet and operational parameters while considering predictability.

## 6. Conclusions

It is self-evident that a DRT service to provide better performance for the user should ensure lower detour times. The implementation of the proposed methodological approach allowed us to quantify the predictability of the DRT service as operating parameters vary, being able to obtain a higher predictability in the case study of Palermo even though setting higher detour times than those of Acireale. In fact, to obtain adequate predictability in the case study of Acireale, it was necessary to set higher performance service characteristics (i.e. lower detour times) compared to those established for Palermo. Thus, this approach served to evaluate not only the predictability associated with the DRT service, but also constituted an operational tool to support the choice of parameters for the operational design of the service. Additionally, by integrating spatial and temporal demand variability into our framework, we provide a more accurate tool for evaluating the predictability of on-demand transport services.

More generally, the research findings demonstrate that for services characterized by great flexibility allow a good predictability satisfying more requests. Even more, we show that the level of predictability of DRT can be seen as a design parameter of the service and can be decided at a strategic level, during system dimensioning. This allows to optimize the service reducing costs and maintaining high predictability. Future insights will include an analysis of the costs associated with the service and the application to different case studies. Reliable detour estimation is crucial to assess accessibility prompted by implementing these services.

**Ackowledgements**

This work has been supported by the French ANR research project MuTAS (ANR-21-CE22-0025-01). The work of [XX] is funded by the European Union (Next-Generation EU), through the MUR-PNRR project SAMOTHRACE (ECS00000022).